\documentclass[pra,twocolumn,superscriptaddress,showpacs,nofootinbibfloatfix,amsmath,amsfonts,amssymb]{revtex4-1}%
\usepackage{amsmath,amsfonts,amssymb,color}
\usepackage{amsthm}
\usepackage{leftidx}
\usepackage{graphicx}
\usepackage{xcolor}
\usepackage{dcolumn}
\usepackage{bm}
\usepackage{epstopdf}
\usepackage{epsfig}
\usepackage{environ}
\usepackage{pdfcomment}

\usepackage{float}
\usepackage[T1]{fontenc}
\usepackage[latin9]{inputenc}
\usepackage{setspace}
\usepackage{esint}

\begin{document}

\preprint{APS/123-QED}

\title{Floquet higher-order topological phases in momentum space}

\author{Longwen Zhou}
\email{zhoulw13@u.nus.edu}
\affiliation{%
	Department of Physics, College of Information Science and Engineering, Ocean University of China, Qingdao 266100, China
}

\date{\today}

\begin{abstract}
Higher-order topological phases (HOTPs) are characterized by symmetry-protected
bound states at the corners or hinges of the system. In this work,
we reveal a momentum-space counterpart of HOTPs in time-periodic driven
systems, which are demonstrated in a two-dimensional extension of
the quantum double-kicked rotor. The found Floquet HOTPs are protected
by chiral symmetry and characterized by a pair of topological invariants,
which could take arbitrarily large integer values with the increase
of kicking strengths. These topological numbers can also be measured
from the chiral dynamics of wave packets. Under open boundary conditions,
multiple quartets Floquet corner modes with zero and $\pi$ quasienergies
emerge in the system and coexist with delocalized bulk states at the
same quasienergies, forming second-order Floquet topological bound
states in continuum. The numbers of these corner modes are further
counted by the bulk topological invariants according to the relation
of bulk-corner correspondence. Our findings thus extend the study
HOTPs to momentum-space lattices, and further uncover the richness
of HOTPs and corner-localized bound states in continuum in Floquet
driven systems.
\end{abstract}

\maketitle

\section{Introduction\label{sec:Int}}

Higher-order topological phases (HOTPs) in $D$ spatial dimensions
are characterized by symmetry-protected states localized along its
$(D-n)$-dimensional boundaries, where $1<n\leq D$~\cite{HOTP0,HOTP1,HOTP2,HOTP3,HOTP4,HOTP5,HOTP6,HOTP7}. The existence of
these unique topological matter is usually guaranteed by the coexistence
of crystal and non-spatial symmetries, and their classifications go
beyond the tenfold way of first-order topological insulators and superconductors~\cite{Classify1,Classify2,Classify3,Classify4}.
Besides great theoretical efforts in the study of higher-order topological insulators~\cite{HOTI1,HOTI2,HOTI3,HOTI4,HOTI5,HOTI6,HOTI7,HOTI8,HOTI9,HOTI10,HOTI11,HOTI12,HOTI13,HOTI14,HOTI15,HOTI16,HOTI17,HOTI18,HOTI19,HOTI20}, superconductors~\cite{HOTSC1,HOTSC2,HOTSC3,HOTSC4,HOTSC5,HOTSC6,HOTSC7,HOTSC8,HOTSC9,HOTSC10,HOTSC11,HOTSC12,HOTSC13,HOTSC14,HOTSC15,HOTSC16} and semimetals~\cite{HOTSM1,HOTSM2,HOTSM3,HOTSM4,HOTSM5,HOTSM6}, HOTPs have also been observed in
solid state materials~\cite{SoliStat1,SoliStat2,SoliStat3,SoliStat4,SoliStat5}, photonic waveguides~\cite{Photonic1,Photonic2,Photonic3,Photonic4,Photonic5,Photonic6,Photonic7,Photonic8}, acoustic
systems~\cite{Acoustic1,Acoustic2,Acoustic3,Acoustic4,Acoustic5,Acoustic6,Acoustic7,Acoustic8}, electrical circuits~\cite{Circuit1,Circuit2,Circuit3,Circuit4,Circuit5} and superconducting qubits~\cite{SCQubit1}, leading to potential applications
such as acoustic sensing~\cite{Acoustic1,Acoustic2} and holonomic quantum computation~\cite{HOTSC16}.

In recent years, the study of HOTPs has been generalized to nonequilibrium
systems, such as those subject to time-periodic drivings~\cite{FHOTP1,FHOTP2,FHOTP3,FHOTP4,FHOTP5,FHOTP6,FHOTP7,FHOTP8,FHOTP9,FHOTP10,FHOTP11,FHOTP12,FHOTP13} or non-Hermitian
effects~\cite{NHHOTP1,NHHOTP2,NHHOTP3,NHHOTP4,NHHOTP5,NHHOTP6,NHHOTP7,NHHOTP8,NHHOTP9}. The motivation behind the exploration of HOTPs in periodically
driven systems is threefold. First, driving fields could induce symmetries
and phase transitions that are unique to Floquet systems, yielding
Floquet HOTPs with topological properties that go beyond any static
counterparts~\cite{FHOTP1,FHOTP12}. Second, periodic driving fields could in general enlarge
the range of hoppings in a lattice, creating Floquet HOTPs with large
topological numbers and many topological corner/hinge modes~\cite{FHOTP1}, which
have potential applications in the construction of topological time
crystals and Floquet quantum computing schemes~\cite{FHOTP2}. Third, under appropriate
conditions, certain Floquet systems could form lattice structures
in momentum space, whose topological properties are of intrinsic dynamical
origins. Intriguing phenomena related to such momentum space topology
including the quantized acceleration as an analogue of the Thouless
pump~\cite{DerekPRL}, and the integer quantum Hall effects from chaos~\cite{TianPRL}. The first two
aspects have led to the discoveries of a series of Floquet HOTPs
in both Hermitian and non-Hermitian systems~\cite{FHOTP1,FHOTP2,FHOTP3,FHOTP4,FHOTP5,FHOTP6,FHOTP7,FHOTP8,FHOTP9,FHOTP10,FHOTP11,FHOTP12,FHOTP13,NHHOTP9}.
However, the momentum-space counterpart of Floquet HOTPs and their
topological characterizations have rarely been explored.

In this manuscript, we investigate a periodically kicked particle
in two spatial dimensions, whose momentum space could form a two-dimensional~(2D)
discrete lattice holding rich Floquet HOTPs. In Sec.~\ref{sec:Mod},
we introduce the Hamiltonian of the system and obtain its Floquet
operator under the quantum resonance condition. Based on the symmetry
analysis of the model, we construct a pair of integer topological
invariants $(w_{0},w_{\pi})$ in Sec.~\ref{sec:WN}, which could fully
characterize the Floquet HOTPs that are protected by the chiral symmetry
of the system. These Floquet HOTPs are further shown to be able to
possess arbitrary large topological numbers with the increase of kicking
strengths. In Sec.~\ref{sec:MCD}, we show that these topological
invariants could be dynamically probed by measuring the time-averaged
mean chiral displacements (MCDs) of wave packets in two-dimension.
Under the open boundary conditions (OBCs), we find many quartets of
Floquet corner modes at zero and $\pi$ quasienergies in Sec.~\ref{sec:Corner}.
The numbers of these corner modes are predicted by the bulk topological
invariants $(w_{0},w_{\pi})$, yielding the bulk-corner correspondence
of Floquet HOTPs in momentum space. Moreover, the zero and $\pi$
Floquet corner modes are found to be embedded in the continuous bulk
bands of delocalized states, forming corner-localized Floquet bound
states in the continuum (BICs) that are originated from higher-order
Floquet topology. We summarize our results and discuss potential
future directions in Sec.~\ref{sec:Sum}.

\section{Model\label{sec:Mod}}
In this section, we introduce a representative driven lattice model,
which could possess rich Floquet HOTPs in momentum space. Our system
can be viewed as a two-dimensional extension of the double kicked
rotor (or lattice) model~\cite{DKR1,DKR2,DKR3,DKR4}, which describes a quantum particle kicked
twice by a periodic potential at different times within
each driving period. The time-dependent Hamiltonian of the system
takes the form
\begin{equation}
\hat{H}=\hat{H}_{0}+\hat{V}\sum_{\ell\in\mathbb{Z}}\delta\left(\frac{t}{T}-\ell\right)+\hat{W}\sum_{\ell\in\mathbb{Z}}\delta\left(\frac{t-\tau}{T}-\ell\right),\label{eq:H}
\end{equation}
where
\begin{equation}
\hat{H}_{0}=\frac{\hat{p}_{x}^{2}+\hat{p}_{y}^{2}}{2},\label{eq:H0}
\end{equation}
\begin{equation}
\hat{V}=\kappa_{1}\cos(\hat{x}+\phi_{x})+\kappa_{3}\cos(\hat{y}+\phi_{y}),\label{eq:V}
\end{equation}
and
\begin{equation}
\hat{W}=\kappa_{2}\cos(\hat{x})+\kappa_{4}\cos(\hat{y}).\label{eq:W}
\end{equation}
Here $(\hat{x},\hat{y})$ and $(\hat{p}_{x},\hat{p}_{y})$ are the
position and momentum operators of the particle along $x$ and $y$
directions. $T$ is the driving period. $\tau\in(0,T)$ controls the
time delay between the two kicks inside a driving period. $\kappa_{1,3}$
and $\kappa_{2,4}$ are kicking strengths of the potentials along
$x$ and $y$ directions. $\phi_{x,y}\in[0,2\pi)$ describe the phase
differences between the kicking potentials applied at $t=\ell T$
and $t=\ell T+\tau$ in the $\ell$'s driving period. The quantities
in Eqs.~(\ref{eq:H0})-(\ref{eq:W}) are all set in dimensionless
units. In experiments, the model Hamiltonian $\hat{H}$ may be realized
in cold atom systems, where the kicking potentials could be implemented
by optical-lattice potentials with relative phase shifts~\cite{DKR1,DKR2}. Within
a given driving period (e.g., from $t=0^{-}$ to $t=T+0^{-}$), the
dynamics of the system is therefore governed by a kick $\hat{V}$
applied at $t=0$, followed by the free evolution $\hat{H}_{0}$ from
$t=0\rightarrow\tau$, a second kick $\hat{W}$ at $t=\tau$, and
finally the free evolution $\hat{H}_{0}$ over a time duration $T-\tau$.
The Floquet operator, which describes the evolution of the system
over such a complete driving period, is then given by
\begin{equation}
\hat{U}=e^{-i\frac{T-\tau}{\hbar}\hat{H}_{0}}e^{-i\frac{T}{\hbar}\hat{W}}e^{-i\frac{\tau}{\hbar}\hat{H}_{0}}e^{-i\frac{T}{\hbar}\hat{V}}.\label{eq:U}
\end{equation}

Due to the periodicity of kicking potentials $\hat{V}$ and
$\hat{W}$ in ${\hat x}$ and ${\hat y}$, the eigenvalues of momentum operators $\hat{p}_{x,y}$
take the forms $\hbar(n_{x,y}+k_{x,y})$, where $n_{x,y}\in\mathbb{Z}$,
and $k_{x,y}\in[0,1)$ are the quasimomenta. A Floquet system with
periodicity in momentum space is achieved by setting $k_{x,y}=0$,
which maybe realized experimentally by a Bose-Einstein condensate
with large coherence width~\cite{DKR5,DKR6,DKR7,DKR8}. Under this condition, we can identify
the momentum operators $\hat{p}_{x,y}$ in Eq.~(\ref{eq:H0}) as $\hbar\hat{n}_{x,y}$
with integer eigenvalues $n_{x,y}$. The Floquet operator in Eq.~(\ref{eq:U})
then takes the explicit form
\begin{alignat}{1}
\hat{U}& = e^{-i\hbar(T-\tau)\frac{\hat{n}_{x}^{2}+\hat{n}_{y}^{2}}{2}}e^{-i[K_{2}\cos(\hat{x})+K_{4}\cos(\hat{y})]}\nonumber \\
& \times e^{-i\hbar\tau\frac{\hat{n}_{x}^{2}+\hat{n}_{y}^{2}}{2}}e^{-i[K_{1}\cos(\hat{x}+\phi_{x})+K_{3}\cos(\hat{y}+\phi_{y})]},\label{eq:UDKL}
\end{alignat}
where we have introduced $K_{j}=\kappa_{j}T/\hbar$ as rescaled dimensionless
kicking strengths. Furthermore, under the quantum resonance condition
$\hbar T=4\pi$ that has been considered experimentally~\cite{DKR5,DKR6,DKR7,DKR8,DKR9,DKR10,DKR11}, we obtain
the two-dimensional extension of on-resonance double kicked rotor
(or lattice) model, whose Floquet operator reads
\begin{alignat}{1}
\hat{U}& = e^{i\hbar\tau\frac{\hat{n}_{x}^{2}+\hat{n}_{y}^{2}}{2}}e^{-i[K_{2}\cos(\hat{x})+K_{4}\cos(\hat{y})]}\nonumber \\
& \times e^{-i\hbar\tau\frac{\hat{n}_{x}^{2}+\hat{n}_{y}^{2}}{2}}e^{-i[K_{1}\cos(\hat{x}+\phi_{x})+K_{3}\cos(\hat{y}+\phi_{y})]}.\label{eq:ORDKL}
\end{alignat}
It is then clear that once $\hbar\tau=2\pi p/q$, with $p$ and $q$
being coprime integers, the Floquet operator $\hat{U}$ will have
translational symmetries in both $\hat{n}_{x}$ and $\hat{n}_{y}$ with
the common period $q$, i.e., a periodic crystal structure in the
momentum space of the two-dimensional on-resonance double-kicked lattice
(2D-ORDKL). In one-dimensional (1D) descendant models of Eq.~(\ref{eq:ORDKL}),
rich first-order Floquet topological phases have been discovered,
which are characterized by large Chern (winding numbers), multiple
chiral (dispersionless) edge modes and topologically quantized acceleration
in momentum space~\cite{DerekPRL,ORDKR1,ORDKR2,ORDKR3,ORDKR4}. These discoveries further motivate us to explore
HOTPs in the 2D-ORDKL model.

To obtain a minimal version of the 2D-ORDKL with nontrivial higher-order
topology, we choose the time delay between the two kicks to be $\tau=T/4$,
which implies that $\hbar\tau=\pi$. Moreover, fixing the phase differences
at $\phi_{x}=\phi_{y}=\pi/2$, the Floquet operator of the 2D-ORDKL reduces
to
\begin{alignat}{1}
\hat{U}& = e^{i\frac{\pi}{2}(\hat{n}_{x}^{2}+\hat{n}_{y}^{2})}e^{-i[K_{2}\cos(\hat{x})+K_{4}\cos(\hat{y})]}\nonumber \\
& \times e^{-i\frac{\pi}{2}(\hat{n}_{x}^{2}+\hat{n}_{y}^{2})}e^{i[K_{1}\sin(\hat{x})+K_{3}\sin(\hat{y})]}.\label{eq:2DORDKL}
\end{alignat}
Since $[\hat{n}_{x},\hat{y}]=[\hat{n}_{y},\hat{x}]=0$, the 2D-ORDKL
can be viewed as two 1D ORDKL models lying along two different spatial
dimensions, i.e., $\hat{U}=\hat{U}_{x}\otimes\hat{U}_{y}$, where
\begin{equation}
\hat{U}_{x}=e^{i\frac{\pi}{2}\hat{n}_{x}^{2}}e^{-iK_{2}\cos(\hat{x})}e^{-i\frac{\pi}{2}\hat{n}_{x}^{2}}e^{iK_{1}\sin(\hat{x})},\label{eq:Ux}
\end{equation}
\begin{equation}
\hat{U}_{y}=e^{i\frac{\pi}{2}\hat{n}_{y}^{2}}e^{-iK_{4}\cos(\hat{y})}e^{-i\frac{\pi}{2}\hat{n}_{y}^{2}}e^{iK_{3}\sin(\hat{y})}.\label{eq:Uy}
\end{equation}
By solving the Floquet eigenvalue equations $\hat{U}_{x}|\psi_{x}\rangle=e^{-iE_{x}}|\psi_{x}\rangle$
and $\hat{U}_{y}|\psi_{y}\rangle=e^{-iE_{y}}|\psi_{y}\rangle$, we
could obtain the eigenstates $|\psi\rangle=|\psi_{x}\rangle\otimes|\psi_{y}\rangle$
of $\hat{U}$ with eigenphases (quasienergies) $E=(E_{x}+E_{y})$ mod
$2\pi$. This observation immediately allows us to deduce the possible origin
of higher-order topology in the 2D-ORDKL. That is, if $|\psi_{x}\rangle$ and
$|\psi_{y}\rangle$ are edge modes of $\hat{U}_{x}$ and $\hat{U}_{y}$
with eigenphases $(E_{x},E_{y})=(0,0)$ or $(E_{x},E_{y})=(\pi,\pi)$,
they will be coupled to form a corner mode of $\hat{U}$ with eigenphase
$E=E_{x}+E_{y}=0$, i.e., a Floquet zero corner mode. Similarly, if
$|\psi_{x}\rangle$ and $|\psi_{y}\rangle$ are edge modes of $\hat{U}_{x}$
and $\hat{U}_{y}$ with eigenphases $(E_{x},E_{y})=(0,\pi)$ or $(E_{x},E_{y})=(\pi,0)$,
they will be coupled to form a corner mode of $\hat{U}$ with eigenphase
$E=E_x+E_y=\pi$, i.e., a Floquet $\pi$ corner mode. These are the two types
of topological corner modes that could appear in Floquet HOTPs of the 2D-ORDKL,
and their numbers are determined by the numbers of edge modes in the
subsystems described by $\hat{U}_{x}$ and $\hat{U}_{y}$, which are
further determined by the winding numbers of $\hat{U}_{x}$ and $\hat{U}_{y}$
according to the principle of bulk-edge correspondence~\cite{BBC1,BBC2}. In the following
section, we will construct the bulk topological invariants for 
Floquet HOTPs in the 2D-ORDKL based on these analysis, and establish
the topological phase diagram of the system.

\section{Topological invariants and phase diagram\label{sec:WN}}
Since the Floquet operator $\hat{U}$ in Eq.~(\ref{eq:2DORDKL}) possesses
a tensor product structure, its spectrum and eigenstates are known
once the eigenvalue equations $\hat{U}_{j}|\psi_{j}\rangle=e^{-iE_{j}}|\psi_{j}\rangle$
for $j=x,y$ are solved. Inserting the identity operators $\mathbb{I}_{j}=\sum_{j}|n_{j}\rangle\langle n_{j}|$
in the momentum space, and performing Fourier transforms from the
momentum to quasiposition (the conserved quantity due to the translational
symmetry $n_{j}\rightarrow n_{j}+2$ in the momentum lattice) representation,
the Floquet operator $\hat{U}_{j}$ can be expressed in the form of
\begin{equation}
\hat{U}_{j}=\sum_{\theta_{j}}|\theta_{j}\rangle U_{j}(\theta_{j})\langle\theta_{j}|,\label{eq:Uj}
\end{equation}
where $\{|\theta_{j}\rangle\}$ is the eigenbasis of quasiposition
with $\theta_{j}\in[0,2\pi)$, and $j=x,y$. Explicitly, the Floquet
matrices $U_{x}(\theta_{x})$ and $U_{y}(\theta_{y})$ are given by
\begin{alignat}{1}
U_{x}(\theta_{x})& = e^{i\frac{\pi}{4}\sigma_{z}}e^{-i{\cal K}_{2}\left(\cos\frac{\theta_{x}}{2}\sigma_{x}+\sin\frac{\theta_{x}}{2}\sigma_{y}\right)}\nonumber \\
& \times e^{-i\frac{\pi}{4}\sigma_{z}}e^{i{\cal K}_{1}\left(\cos\frac{\theta_{x}}{2}\sigma_{x}+\sin\frac{\theta_{x}}{2}\sigma_{y}\right)},\label{eq:UxThx}
\end{alignat}
\begin{alignat}{1}
U_{y}(\theta_{y})& = e^{i\frac{\pi}{4}\tau_{z}}e^{-i{\cal K}_{4}\left(\cos\frac{\theta_{y}}{2}\tau_{x}+\sin\frac{\theta_{y}}{2}\tau_{y}\right)}\nonumber \\
& \times e^{-i\frac{\pi}{4}\tau_{z}}e^{i{\cal K}_{3}\left(\cos\frac{\theta_{y}}{2}\tau_{x}+\sin\frac{\theta_{y}}{2}\tau_{y}\right)},\label{eq:UyThy}
\end{alignat}
with shorthand notations
\begin{equation}
{\cal K}_{1}\equiv K_{1}\sin\frac{\theta_{x}}{2},\qquad{\cal K}_{2}\equiv K_{2}\cos\frac{\theta_{x}}{2},\label{eq:K12}
\end{equation}
\begin{equation}
{\cal K}_{3}\equiv K_{3}\sin\frac{\theta_{y}}{2},\qquad{\cal K}_{4}\equiv K_{4}\cos\frac{\theta_{y}}{2}.\label{eq:K34}
\end{equation}
Here $\sigma_{x,y,z}$ and $\tau_{x,y,z}$ denote Pauli matrices acting
on two sublattice degrees of freedom along the $x$ and $y$ directions
in the momentum lattice (see Refs.~\cite{ORDKR2,ORDKR3} for derivation details
of $U_{j}(\theta_{j})$ for 1D descendant models of the
2D-ORDKL). The standard characterization of 1D Floquet
topological phases is achieved by introducing a pair of symmetric
time frames upon similarity transformations~\cite{BBC1,BBC2}. For our 2D-ORDKL,
there are two symmetric time frames for both $U_{x}(\theta_{x})$
and $U_{y}(\theta_{y})$. Putting together, there are in total four
such time frames for the Floquet matrix $U(\theta_{x},\theta_{y})=U_{x}(\theta_{x})\otimes U_{y}(\theta_{y})$
of the 2D system. In these time frames, $U(\theta_{x},\theta_{y})$
takes the form
\begin{equation}
U_{\alpha\beta}(\theta_{x},\theta_{y})=U_{\alpha}(\theta_{x})\otimes U_{\beta}(\theta_{y}),\label{eq:Uab}
\end{equation}
where $\alpha=1,2$, $\beta=3,4$, and 
\begin{equation}
U_{1}(\theta_{x})=FG,\qquad U_{2}(\theta_{x})=GF,\label{eq:U12}
\end{equation}
\begin{equation}
U_{3}(\theta_{y})=F'G',\qquad U_{4}(\theta_{y})=G'F'.\label{eq:U34}
\end{equation}
The auxiliary matrices $F$, $G$, $F'$ and $G'$ are explicitly
given by (see Refs.~\cite{ORDKR2,ORDKR3} for derivation details of these
matrices for 1D descendant models of the 2D-ORDKL)
\begin{equation}
F\equiv e^{i\frac{{\cal K}_{1}}{2}\left(\cos\frac{\theta_{x}}{2}\sigma_{x}+\sin\frac{\theta_{x}}{2}\sigma_{y}\right)}e^{i\frac{\pi}{4}\sigma_{z}}e^{-i\frac{{\cal K}_{2}}{2}\left(\cos\frac{\theta_{x}}{2}\sigma_{x}+\sin\frac{\theta_{x}}{2}\sigma_{y}\right)},\label{eq:F}
\end{equation}
\begin{equation}
G\equiv e^{-i\frac{{\cal K}_{2}}{2}\left(\cos\frac{\theta_{x}}{2}\sigma_{x}+\sin\frac{\theta_{x}}{2}\sigma_{y}\right)}e^{-i\frac{\pi}{4}\sigma_{z}}e^{i\frac{{\cal K}_{1}}{2}\left(\cos\frac{\theta_{x}}{2}\sigma_{x}+\sin\frac{\theta_{x}}{2}\sigma_{y}\right)},\label{eq:G}
\end{equation}
\begin{equation}
F'\equiv e^{i\frac{{\cal K}_{3}}{2}\left(\cos\frac{\theta_{y}}{2}\tau_{x}+\sin\frac{\theta_{y}}{2}\tau_{y}\right)}e^{i\frac{\pi}{4}\tau_{z}}e^{-i\frac{{\cal K}_{4}}{2}\left(\cos\frac{\theta_{y}}{2}\tau_{x}+\sin\frac{\theta_{y}}{2}\tau_{y}\right)},\label{eq:F'}
\end{equation}
\begin{equation}
G'\equiv e^{-i\frac{{\cal K}_{4}}{2}\left(\cos\frac{\theta_{y}}{2}\tau_{x}+\sin\frac{\theta_{y}}{2}\tau_{y}\right)}e^{-i\frac{\pi}{4}\tau_{z}}e^{i\frac{{\cal K}_{3}}{2}\left(\cos\frac{\theta_{y}}{2}\tau_{x}+\sin\frac{\theta_{y}}{2}\tau_{y}\right)}.\label{eq:G'}
\end{equation}
With these considerations, it is straightforward to verify that $U_{\alpha\beta}(\theta_{x},\theta_{y})$
possesses the chiral symmetry $\Gamma=\sigma_{z}\otimes\tau_{z}$
for all $\alpha=1,2$ and $\beta=3,4$, in the sense that
\begin{equation}
\Gamma U_{\alpha\beta}(\theta_{x},\theta_{y})\Gamma=U_{\alpha\beta}^{\dagger}(\theta_{x},\theta_{y}).\label{eq:CS}
\end{equation}
This symmetry then allows us to characterize the Floquet HOTPs of the 2D-ORDKL by integer
topological invariants~\cite{ORDKR2,BBC1,BBC2}.

To construct these topological numbers for our system, we take the
Taylor expansion for each term of the Floquet matrices in Eqs.~(\ref{eq:U12}) and
(\ref{eq:U34}), and recombine the relevant terms. The resulting Floquet
matrices take the forms
\begin{equation}
U_{\alpha}(\theta_{x})=e^{-iE_{x}(\theta_{x})\left[n_{\alpha x}(\theta_{x})\sigma_{x}+n_{\alpha y}(\theta_{x})\sigma_{y}\right]},\label{eq:UaThx}
\end{equation}
\begin{equation}
U_{\beta}(\theta_{y})=e^{-iE_{y}(\theta_{y})\left[n_{\beta x}(\theta_{y})\tau_{x}+n_{\beta y}(\theta_{y})\tau_{y}\right]},\label{eq:UbThy}
\end{equation}
where $\alpha=1,2$ and $\beta=3,4$. The eigenphase dispersions $E_{x}(\theta_{x})$
and $E_{y}(\theta_{y})$ along the two different dimensions are given
by
\begin{equation}
E_{x}(\theta_{x})=\arccos\left(\cos{\cal K}_{1}\cos{\cal K}_{2}\right),\label{eq:ExThx}
\end{equation}
\begin{equation}
E_{y}(\theta_{y})=\arccos\left(\cos{\cal K}_{3}\cos{\cal K}_{4}\right).\label{eq:EyThy}
\end{equation}
The explicit expressions of unit vectors $[n_{\alpha x}(\theta_{x}),n_{\alpha y}(\theta_{x})]$
and $[n_{\beta x}(\theta_{y}),n_{\beta y}(\theta_{y})]$ are summarized
in the Appendix \ref{sec:AppA}. In the quasiposition representation,
the 2D-ORDKL model then possesses four eigenphase (quasienergy) bands,
whose dispersion relations are given by
\begin{equation}
E_{ss'}(\theta_{x},\theta_{y})=sE_{x}(\theta_{x})+s'E_{y}(\theta_{y}),\label{eq:EThxThy}
\end{equation}
where $s,s'=\pm$. The system could undergo topological phase transitions when these bands touch and separate at the quasienergies zero and $\pi$.

In previous studies~\cite{ORDKR2,ORDKR3}, it has been demonstrated that a 1D
system described by the Floquet operator $U_{\alpha}(\theta_{x})$
or $U_{\beta}(\theta_{y})$ possesses a topological winding number
\begin{equation}
w_{\nu}=\int_{0}^{2\pi}\frac{d\theta_{j}}{2\pi}\partial_{\theta_{j}}\varphi_{\nu}(\theta_{j}),\label{eq:WN}
\end{equation}
where $\nu=\{\alpha,\beta\}$, $j=x,y$, and the winding angle $\varphi_{\nu}(\theta_{j})$
in the $\nu$'s time frame is defined as
\begin{equation}
\varphi_{\nu}(\theta_{j})\equiv\arctan\left[n_{\nu y}(\theta_{j})/n_{\nu x}(\theta_{j})\right].\label{eq:PhinThj}
\end{equation}
Using these winding numbers, we can further construct two pairs of invariants
$(w_{0x},w_{\pi x})$ and $(w_{0y},w_{\pi y})$ for the Floquet subsystems
described by $\hat{U}_{x}$ and $\hat{U}_{y}$, respectively~\cite{BBC1,BBC2}. They
are related to the values of $w_{\nu}$ through the relations
\begin{equation}
w_{0x}=\frac{w_{1}+w_{2}}{2},\qquad w_{\pi x}=\frac{w_{1}-w_{2}}{2},\label{eq:W0PX}
\end{equation}
\begin{equation}
w_{0y}=\frac{w_{3}+w_{4}}{2},\qquad w_{\pi y}=\frac{w_{3}-w_{4}}{2}.\label{eq:W0PY}
\end{equation}
These invariants always take integer quantized values. They
provide a complete characterization of the topological phases in 1D
Floquet systems with chiral (sublattice) symmetry~\cite{BBC1,BBC2}. Moreover,
under the open boundary condition, the invariants $w_{0x}$ ($w_{0y}$)
and $w_{\pi x}$ ($w_{\pi y}$) could predict the numbers of topological
edge modes with quasienergies zero and $\pi$ in the subsystem described
by $\hat{U}_{x}$ ($\hat{U}_{y}$)~\cite{ORDKR2,ORDKR3},
and therefore also capturing the bulk-edge correspondence of these Floquet subsystems.

For our 2D-ORDKL model, the Floquet HOTPs can be characterized by
appropriate combinations of these 1D topological numbers.
Specially, referring to our analysis on how the zero and $\pi$ Floquet
edge modes can be coupled to form corner modes in the last section,
we introduce a pair of topological invariants $(w_{0},w_{\pi})$ for
the 2D-ORDKL, which are defined as
\begin{equation}
w_{0}\equiv|w_{0x}w_{0y}|+|w_{\pi x}w_{\pi y}|,\label{eq:W0}
\end{equation}
\begin{equation}
w_{\pi}\equiv|w_{0x}w_{\pi y}|+|w_{\pi x}w_{0y}|.\label{eq:WP}
\end{equation}
It is clear that $(w_{0},w_{\pi})\in\mathbb{Z}\times\mathbb{Z}$ due
to the quantized nature of $w_{0j}$ and $w_{\pi j}$ ($j=x,y$).
Furthermore, as will be demonstrated in Sec.~\ref{sec:Corner}, the
values of $w_{0}$ and $w_{\pi}$ could correctly count the numbers
of Floquet corner modes with quasienergies zero and $\pi$ in the
momentum space of the 2D-ORDKL. Therefore, the invariants $(w_{0},w_{\pi})$
could provide us with a complete characterization of Floquet HOTPs
in the 2D-ORDKL, and other chiral symmetric 2D lattice models with
Floquet operators in the form of $\hat{U}=\hat{U}_{x}\otimes\hat{U}_{y}$.
We also notice that $(w_{0},w_{\pi})\neq(0,0)$ only when the subsystems described by $\hat{U}_{x}$ and $\hat{U}_{y}$ are both topologically
nontrivial. The Floquet HOTPs of the 2D-ORDKL are thus originated
from the cooperation of topological natures of the two subsystems in lower dimensions.

\begin{figure}
	\begin{centering}
		\includegraphics[scale=0.49]{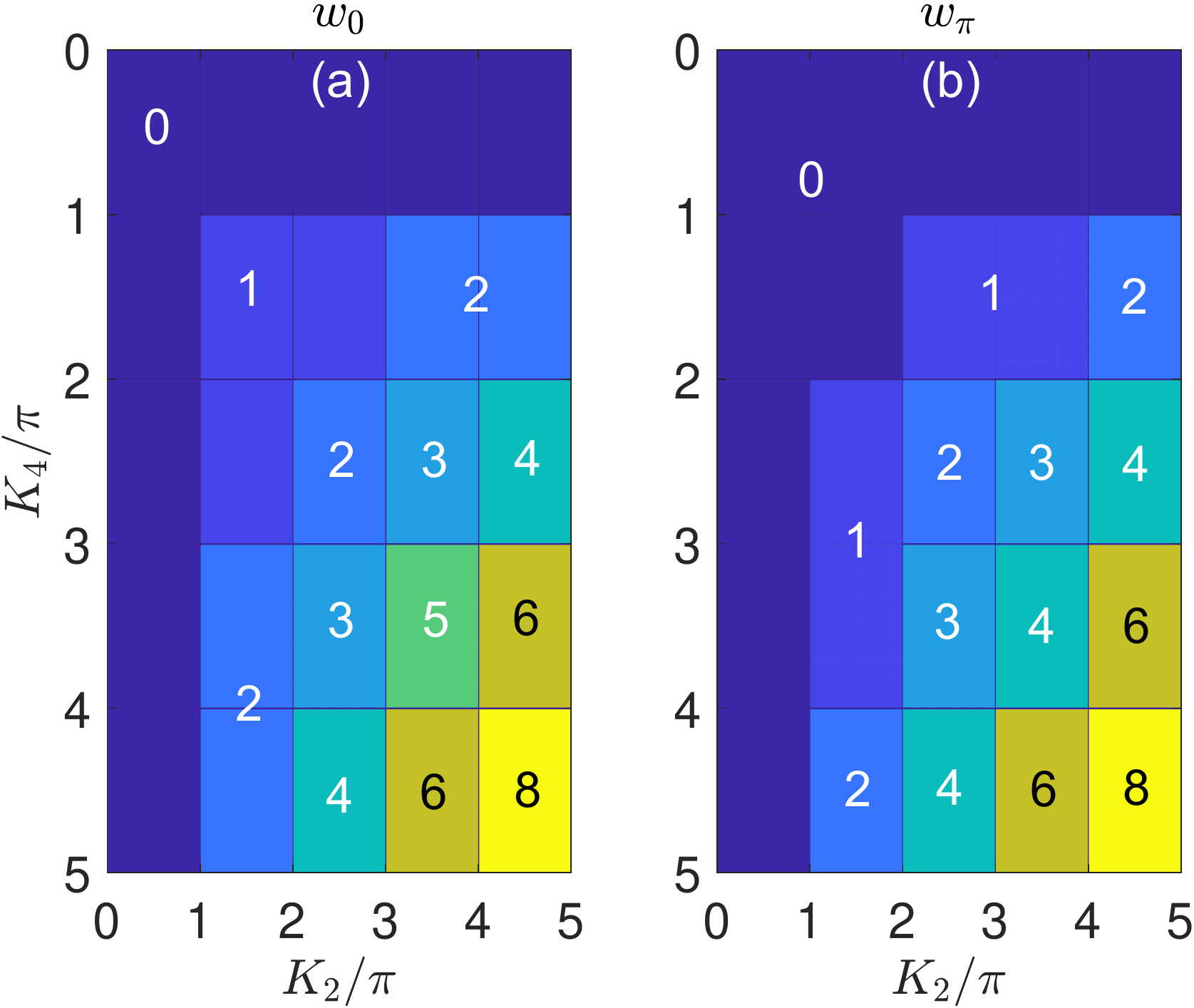}
		\par\end{centering}
	\caption{Topological phase diagram of the 2D-ORDKL versus the kicking strengths
		$K_{2}$ and $K_{4}$. The other system parameters are set as $K_{1}=K_{3}=0.5\pi$.
		Different Floquet HOTPs are distinguished by different colors in each
		figure panel. The values of $w_{0}$ ($w_{\pi}$) for different topological
		phases are obtained from Eq.~(\ref{eq:W0}) {[}(\ref{eq:WP}){]} and
		denoted explicitly in panel (a) {[}(b){]}.\label{fig:WN1}}
\end{figure}

In the remaining part of this section, based on the evaluation
of invariants $w_{0}$ and $w_{\pi}$ in Eqs.~(\ref{eq:W0}) and (\ref{eq:WP}),
we present topological phase
diagrams of the 2D-ORDKL for two typical situations.
In the first case, we show the phase diagram of the system with
respect to the kicking strengths $(K_{2},K_{4})$ in Fig.~\ref{fig:WN1}.
We observe that with the increase of these kicking strengths, a series
of topological phase transitions can be induced, which each of them
being accompanied by the quantized jump of $w_{0}$ or $w_{\pi}$.
At large values of $(K_{2},K_{4})$, we further obtain rich Floquet
HOTPs characterized by large values of $(w_{0},w_{\pi})$. It is not
hard to verify that there is no upper bound for the values of these
invariants if the kicking strengths keep increasing. Therefore, the
2D-ORDKL serves as a good candidate to generate Floquet HOTPs in momentum
space with arbitrarily large topological invariants. Besides, according
to the tensor product structure of Floquet operator in Eq.~(\ref{eq:Uab}),
the boundaries separating different Floquet HOTPs are determined by
the gapless conditions along one dimension of the lattice. More
explicitly, these phase boundaries are determined by the conditions $\cos[E_{x}(\theta_{x})]=\pm1$
and $\cos[E_{y}(\theta_{y})]=\pm1$, or equivalently
\begin{equation}
\frac{m_{1}^{2}}{K_{1}^{2}}+\frac{m_{2}^{2}}{K_{2}^{2}}=\frac{1}{\pi^{2}},\qquad\frac{m_{3}^{2}}{K_{3}^{2}}+\frac{m_{4}^{2}}{K_{4}^{2}}=\frac{1}{\pi^{2}},\label{eq:G0P}
\end{equation}
with $m_{i}\in\mathbb{Z}$ and $|m_{i}\pi/K_{i}|\leq1$ for $i=1,2,3,4$.
In Fig.~\ref{fig:WN1}, our choice of system parameters yield $m_{1}=m_{3}=0$,
and the phase boundaries are reduced to straight lines according to Eq.~(\ref{eq:G0P}), which is also consistent with the numerical results. 

\begin{figure}
	\begin{centering}
		\includegraphics[scale=0.49]{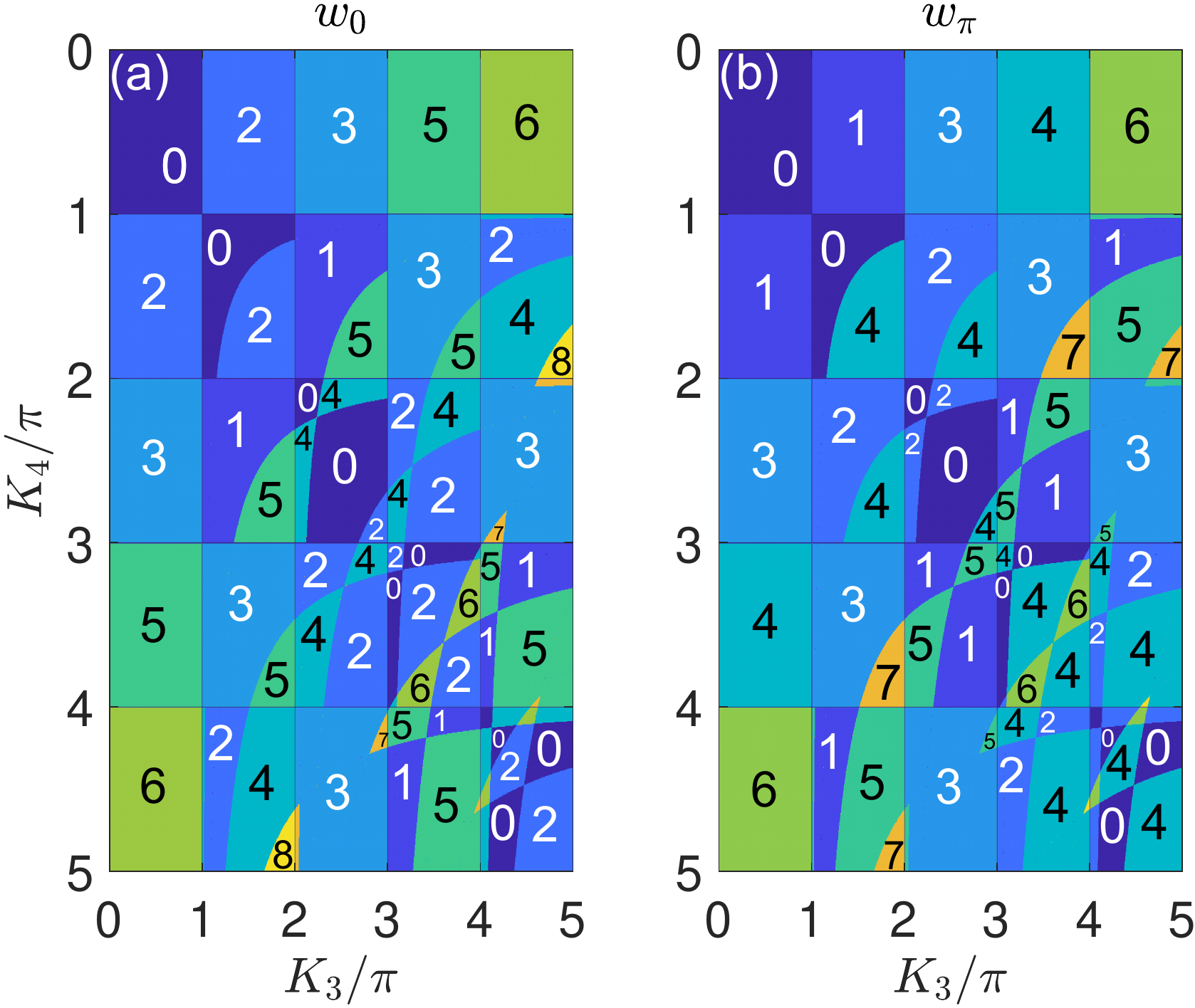}
		\par\end{centering}
	\caption{Topological phase diagram of the 2D-ORDKL with respect to the kicking
		strengths $K_{3}$ and $K_{4}$. The other system parameters are set
		as $(K_{1},K_{2})=(0.5\pi,3.5\pi)$. Different Floquet HOTPs are distinguished
		by different colors in both figure panels. The values of $w_{0}$
		and $w_{\pi}$ for each topological phase are obtained from Eqs.~(\ref{eq:W0})
		and (\ref{eq:WP}), which are presented explicitly in figure panels
		(a) and (b), respectively.\label{fig:WN2}}
\end{figure}

In the second case, we present the phase diagram of the 2D-ORDKL versus
kicking strengths $(K_{3},K_{4})$ in Fig.~\ref{fig:WN2}. With the
increase of these kicking strengths, we also observe rich Floquet
HOTPs featured by large invariants $(w_{0},w_{\pi})$, together with
multiple transitions between these phases followed by quantized changes
of $(w_{0},w_{\pi})$. Numerically, we have also checked the phase
diagrams of the system versus any one pair of kicking parameters
$(K_{i},K_{i'})$, with the other pair being fixed for $i,i'=1,2,3,4$,
and obtain similar patterns for the topological phases and phase transitions.
Therefore, we conclude that the 2D-ORDKL indeed possesses rich Floquet
HOTPs, which are characterized by a pair integer topological invariants
$(w_{0},w_{\pi})$. These invariants could not only predict the numbers
of Floquet zero and $\pi$ corner modes in the system under OBCs, but also be detectable experimentally from the dynamics
of wavepackets, as will be presented in the following sections.

\section{Mean chiral displacements\label{sec:MCD}}

In this section, we sketch a dynamical approach that can be used to
probe the invariants $(w_{0},w_{\pi})$ of Floquet HOTPs in the 2D-ORDKL.
This approach is based on the detection of a 2D extension of the time-averaged
mean chiral displacement (MCD), which was introduced in Refs.~\cite{FHOTP1,NHHOTP9}.
In a 2D lattice~(either in position or momentum space), we define
the chiral displacement operator of the dynamical evolution in the time frame $(\alpha,\beta)$
as
\begin{equation}
\hat{C}_{\alpha\beta}(t)=\hat{U}_{\alpha\beta}^{-t}(\hat{N}_{x}\otimes\Gamma_{x})\otimes(\hat{N}_{y}\otimes\Gamma_{y})\hat{U}_{\alpha\beta}^{t}.\label{eq:Cabt}
\end{equation}
Here $t$ counts the number of evolution periods. The Floquet operator
$\hat{U}_{\alpha\beta}=\hat{U}_{\alpha}\otimes\hat{U}_{\beta}$, as
defined in Eq.~(\ref{eq:Uab}) for $\alpha=1,2$ and $\beta=3,4$.
$\hat{N}_{x}$ and $\hat{N}_{y}$ are unit-cell position operators
along $x$ and $y$ directions of the lattice. $\Gamma_{x}$ and $\Gamma_{y}$
describe chiral symmetries of the descendant systems $\hat{U}_{\alpha}$
and $\hat{U}_{\beta}$ along $x$ and $y$ directions, respectively.
For the 2D-ORDKL model, we have $\Gamma_{x}=\sigma_{z}$ and $\Gamma_{y}=\tau_{z}$,
whose tensor product gives the chiral symmetry operator $\Gamma$
of the 2D system. 

To extract the topological winding numbers of $\hat{U}_{\alpha\beta}$,
we initialize the system in a fully polarized state at the central
unit cell of the lattice. The initial state vector then takes the
form
\begin{equation}
|\psi_{0}\rangle=|0_{x}\rangle\otimes|\uparrow_{x}\rangle\otimes|0_{y}\rangle\otimes|\uparrow_{y}\rangle,\label{eq:Psi0}
\end{equation}
where $|0_{x}\rangle$ ($\text{|\ensuremath{0_{y}\rangle}}$) is the
eigenstate of $\hat{N}_{x}$ ($\hat{N}_{y}$) with eigenvalue $N_{x}=0$
($N_{y}=0$). $|\uparrow_{x}\rangle$ and $|\uparrow_{y}\rangle$
are the eigenvectors of $\Gamma_{x}$ and $\Gamma_{y}$ with eigenvalues
$+1$. After the evolution over a number of $t$'s driving periods
in the time frame $(\alpha,\beta)$, the MCD of initial state $|\psi_{0}\rangle$
is given by the expectation value
\begin{equation}
C_{\alpha\beta}(t)=\langle\psi_{0}|\hat{C}_{\alpha\beta}(t)|\psi_{0}\rangle.\label{eq:MCDab}
\end{equation}
Since $|\psi_{0}\rangle$ is not an eigenstate of $\hat{U}_{\alpha\beta}$,
$C_{\alpha\beta}(t)$ is expected to be an oscillating function of
time. To extract the topological information from $C_{\alpha\beta}(t)$,
we take the average $C_{\alpha\beta}(t)$ over many driving periods,
which in the long-time limit yields the stroboscopic time-averaged MCD
\begin{equation}
\overline{C}_{\alpha\beta}=\lim_{t\rightarrow\infty}\frac{1}{t}\sum_{t'=1}^{t}C_{\alpha\beta}(t').\label{eq:MMCDab}
\end{equation}
Following the derivation steps as detailed in Ref.~\cite{FHOTP1}, it can be
shown that for $\alpha=1,2$ and $\beta=3,4$,
\begin{equation}
\overline{C}_{\alpha\beta}=\frac{w_{\alpha}w_{\beta}}{4}.\label{eq:MCDW}
\end{equation}
Here $w_{\alpha}$ and $w_{\beta}$ are the winding numbers defined
in Eq.~(\ref{eq:WN}). With the help of Eqs.~(\ref{eq:W0PX}) and
(\ref{eq:W0PY}), we can recombine the time-averaged MCDs to obtain
the products of winding numbers $w_{0x}w_{0y}$, $w_{0x}w_{\pi y}$,
$w_{\pi x}w_{0y}$ and $w_{\pi x}w_{\pi y}$, which finally yield
the invariants $(w_{0},w_{\pi})$ of the Floquet HOTPs. From now on,
we will denote the recombined time-averaged MCDs that are related
to $w_{0}$ and $w_{\pi}$ as $\overline{C}_{0}$ and $\overline{C}_{\pi}$,
respectively (see Appendix~\ref{sec:AppB} for their explicit expressions).

\begin{figure}
	\begin{centering}
		\includegraphics[scale=0.49]{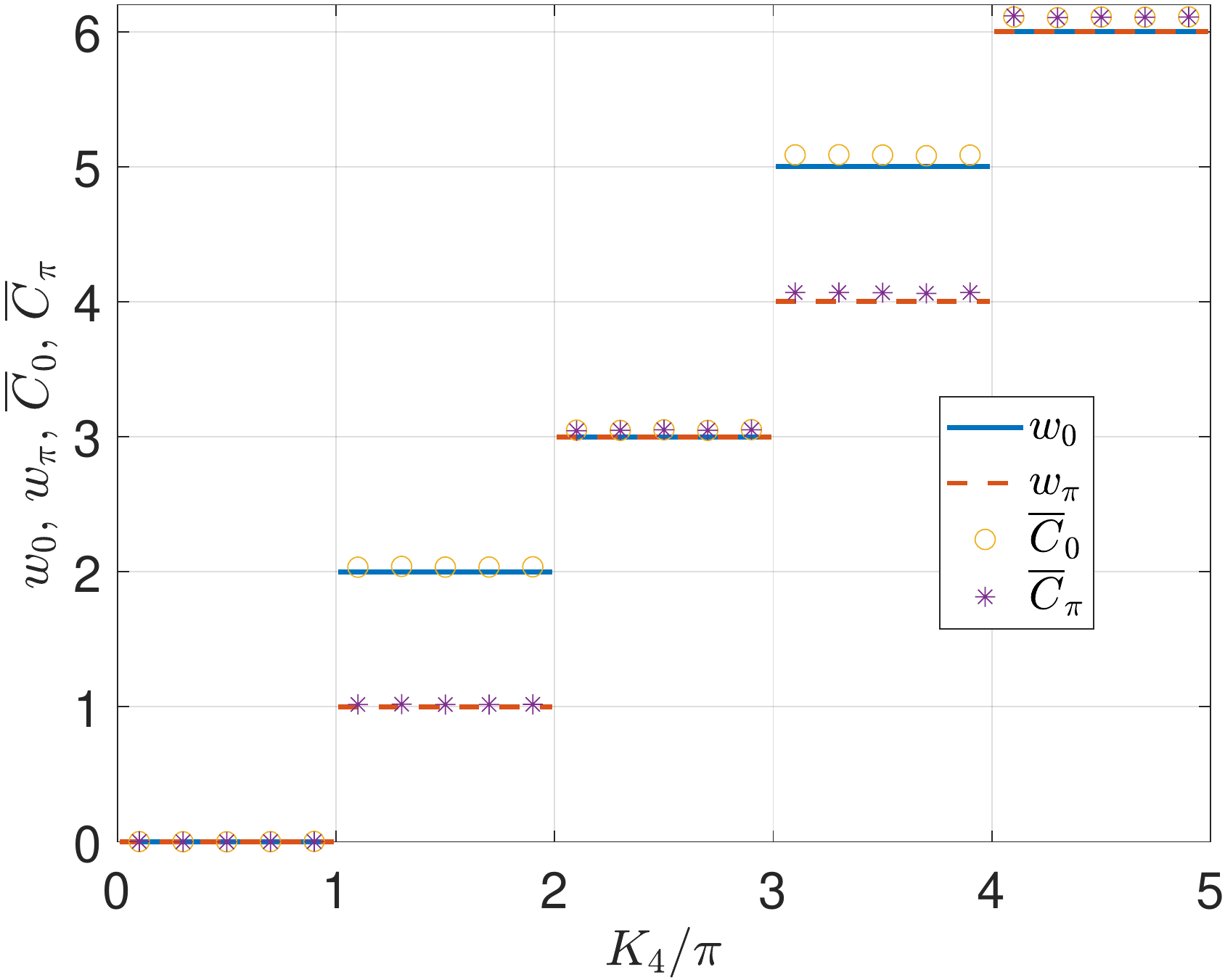}
		\par\end{centering}
	\caption{Topological invariants $(w_{0},w_{\pi})$ and time-averaged MCDs $(\overline{C}_{0},\overline{C}_{\pi})$
		of the 2D-ORDKL versus the kicking strength $K_{4}$. The other system
		parameters are set as $K_{1}=0.5\pi$, $K_{2}=3.5\pi$ and $K_{3}=0.5\pi$.
		The MCDs are averaged over $t=50$ driving periods to generate the
		results for $(\overline{C}_{0},\overline{C}_{\pi})$.\label{fig:MCD}}
\end{figure}

To verify the relations between the time-averaged MCDs and the topological
invariants of the 2D-ORDKL, we compute and compare $(w_{0},w_{\pi})$
and $(\overline{C}_{0},\overline{C}_{\pi})$ for a typical case in
the remaining part of this section. In Fig.~\ref{fig:MCD}, we present
the topological invariants and MCDs with respect to the kicking strength
$K_{4}$ for evolutions over $t=50$ driving periods. We observe that
the time-averaged MCDs $(\overline{C}_{0},\overline{C}_{\pi})$ indeed
take nearly quantized values in each Floquet HOTPs of the 2D-ORDKL,
which demonstrate the relation between them and the topological invariants
of the system, i.e., $(\overline{C}_{0},\overline{C}_{\pi})=(w_{0},w_{\pi})$
(see Appendix \ref{sec:AppB} for derivation details of this relation).
Besides, the values of $(\overline{C}_{0},\overline{C}_{\pi})$ also
possess quantized jumps around all topological phase transition points
($K_{4}=\pi,2\pi,3\pi,4\pi$ in Fig. \ref{fig:MCD}). Therefore,
the MCDs could also serve as a dynamical prob to the phase transitions
between different Floquet HOTPs. The deviations of $(\overline{C}_{0},\overline{C}_{\pi})$
from quantization is due to finite-time effects, which will gradually
go to zero with the increase of the number of driving periods $t$.
Numerically, we have checked that the time-averaged MCDs $(\overline{C}_{0},\overline{C}_{\pi})$
remain close to quantization for $t=20$ driving periods, which should
be within reach in current or near-term experiments in photonic~\cite{MCD1,MCD2}
and cold atom~\cite{MCD3,MCD4} systems.

In the following section, we will demonstrate another topological
signature of Floquet HOTPs, i.e., the symmetry protected bound states
localized around the corners of the momentum-space lattice, and relate
their numbers to the topological invariants $(w_{0},w_{\pi})$ of
the system. 

\section{Floquet topological corner bound states in continuum\label{sec:Corner}}

A defining feature of the HOTP in $D$ spatial dimensions are symmetry-protected
states localized along its $(D-d)$-dimensional boundaries, where $d>1$.
For a 2D lattice studied in this work, such bound states could appear
at the geometric corners of the system. Moreover, since the Floquet
bands of a chiral symmetric system come in positive and negative pairs
$\pm E$, they could touch and separate at the quasienergies zero
and $\pi$. Therefore, in principle there could be two distinct types
of Floquet corner modes appearing at these quasienergies, which will
be called Floquet zero and $\pi$ corner modes. In the 2D-ORDKL described
by $\hat{U}=\hat{U}_{x}\otimes\hat{U}_{y}$, since these corner modes
are formed by the coupling between doubly degenerate edge modes of 1D descendant
systems $\hat{U}_{x}$ and $\hat{U}_{y}$, their numbers $(N_{0},N_{\pi})$
are always integer multiples of four. Furthermore, the numbers $(N_{0},N_{\pi})$
of Floquet zero and $\pi$ corner modes tend out to be connected with
the bulk topological invariants $(w_{0},w_{\pi})$ of the 2D-ORDKL
through the relation
\begin{equation}
(N_{0},N_{\pi})=4(w_{0},w_{\pi}).\label{eq:BCC}
\end{equation}
The Eq.~(\ref{eq:BCC}) thus establishes the bulk-corner correspondence
of Floquet HOTPs in the 2D-ORDKL and other chiral symmetric systems,
whose Floquet operators can be expressed in the tensor product form
of $\hat{U}=\hat{U}_{x}\otimes\hat{U}_{y}$. The implication of Eq.~(\ref{eq:BCC}) will be demonstrated in the following with explicit
numerical examples. 

To investigate the spectrum and states of the system under OBCs, we express the operators in Eqs.~(\ref{eq:Ux}) and (\ref{eq:Uy})
in momentum representations as
\begin{alignat}{1}
\hat{U}_{x}& = e^{i\frac{\pi}{2}\sum_{n_{x}}n_{x}^{2}|n_{x}\rangle\langle n_{x}|}e^{-i\frac{K_{2}}{2}\sum_{n_{x}}(|n_{x}\rangle\langle n_{x}+1|+{\rm h.c.})}\nonumber \\
& \times e^{-i\frac{\pi}{2}\sum_{n_{x}}n_{x}^{2}|n_{x}\rangle\langle n_{x}|}e^{i\frac{K_{1}}{2i}\sum_{n_{x}}(|n_{x}\rangle\langle n_{x}+1|-{\rm h.c.})},\label{eq:UxOBC}
\end{alignat}
\begin{alignat}{1}
\hat{U}_{y}& = e^{i\frac{\pi}{2}\sum_{n_{y}}n_{y}^{2}|n_{y}\rangle\langle n_{y}|}e^{-i\frac{K_{4}}{2}\sum_{n_{y}}(|n_{y}\rangle\langle n_{y}+1|+{\rm h.c.})}\nonumber \\
& \times e^{-i\frac{\pi}{2}\sum_{n_{y}}n_{y}^{2}|n_{y}\rangle\langle n_{y}|}e^{i\frac{K_{3}}{2i}\sum_{n_{y}}(|n_{y}\rangle\langle n_{y}+1|-{\rm h.c.})}.\label{eq:UyOBC}
\end{alignat}
The quasienergies $E$ and Floquet eigenstates $|\psi\rangle$ of
the system are then obtained by solving the eigenvalue equation $\hat{U}|\psi\rangle=e^{-iE}|\psi\rangle$,
where the diagonalization of $\hat{U}$ can be performed separately
for $\hat{U}_{x}$ and $\hat{U}_{y}$ due to the tensor product structure
of the Floquet operator $\hat{U}=\hat{U}_{x}\otimes\hat{U}_{y}$.

\begin{figure}
	\begin{centering}
		\includegraphics[scale=0.49]{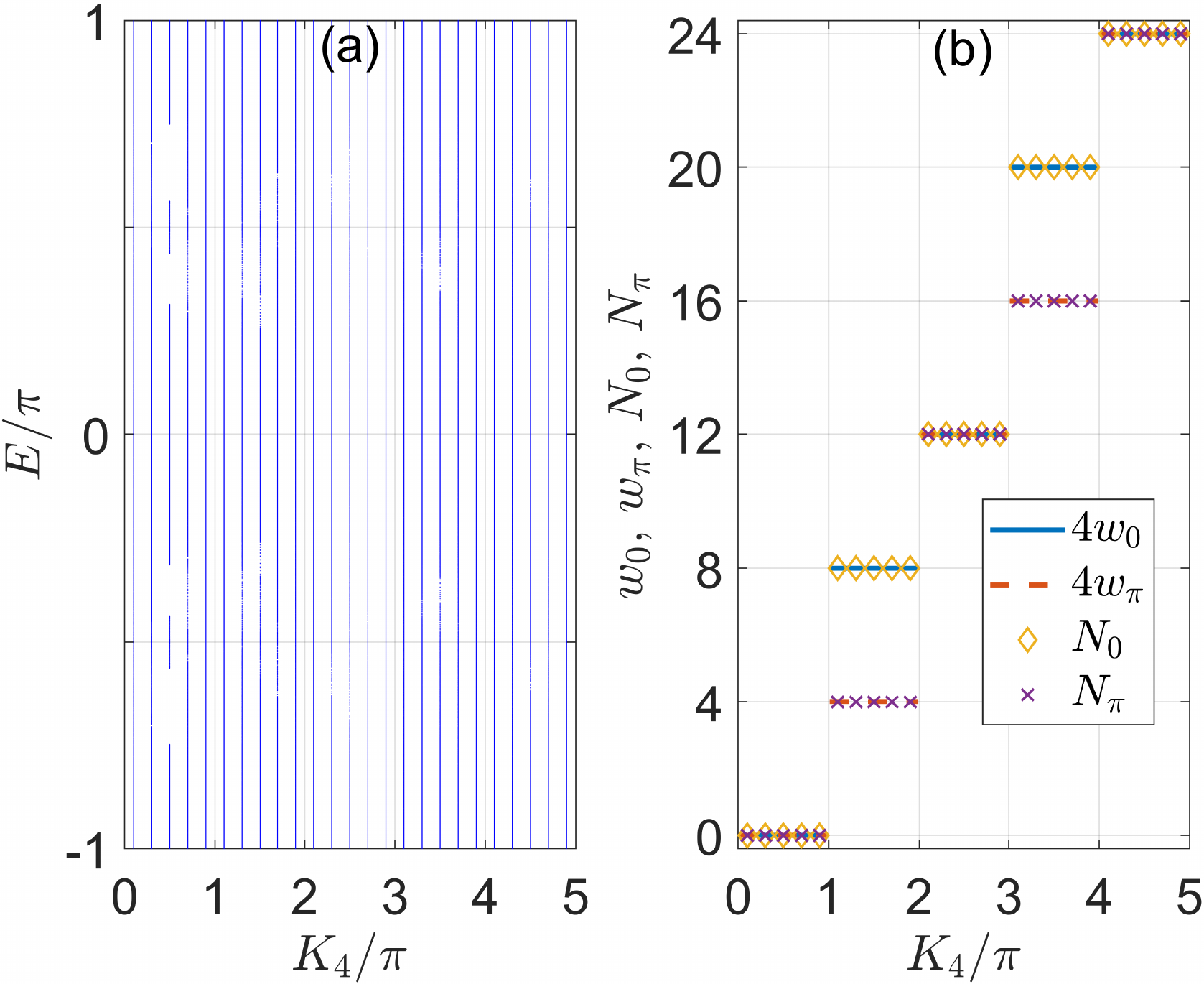}
		\par\end{centering}
	\caption{Floquet spectrum $E$ and number of corner modes $(N_{0},N_{\pi})$
		of the 2D-ORDKL versus the kicking strength $K_{4}$. The other system
		parameters are set as $(K_{1},K_{2},K_{3})=(0.5\pi,3.5\pi,0.5\pi)$.
		The size of the lattice is $L_{x}=L_{y}=300$. In panel (b), $N_{0}$
		and $N_{\pi}$ are plotted together with the bulk topological invariants
		$w_{0}$ and $w_{\pi}$ as obtained from Eqs.~(\ref{eq:W0}) and (\ref{eq:WP}).\label{fig:E}}
\end{figure}

In Fig.~\ref{fig:E}, we present the quasienergy spectrum $E$ and
number of Floquet corner modes $N_{0}$ and $N_{\pi}$ at zero and
$\pi$ quasienergies in the momentum space of 2D-ORDKL. In Fig.~\ref{fig:E}(a),
we observe an almost continuous Floquet spectrum with no gaps around
the quasienergies $E=0$ and $E=\pi$. This is different from the
situations usually observed in 2D static and Floquet HOTPs, where
corner modes are separated from bulk states by spectral gaps.
However, by evaluating the inverse partition ratio~(IPR)
\begin{equation}
{\rm IPR}\equiv\sum_{n_{x},n_{y}}|\psi(n_{x},n_{y})|^{4}\label{eq:IPR}
\end{equation}
for all the Floquet eigenstates of $\hat{U}$, we find different numbers
of corner modes $N_{0}$ and $N_{\pi}$ at the quasienergies zero
and $\pi$ in distinct Floquet HOTPs of the system. Their numbers are presented
together with the topological invariants $w_{0}$ and $w_{\pi}$ in
Fig.~\ref{fig:E}(b). Note that the IPRs of corner modes are differ
from bulk and 1D edge states of the system in their order
of magnitudes, and can thus be numerically distinguished from one
another. Furthermore, we observe that the relation $(N_{0},N_{\pi})=4(w_{0},w_{\pi})$
holds throughout the considered parameter regime, validating the bulk-corner
correspondence of the 2D-ORDKL as established in Eq.~(\ref{eq:BCC}).
Besides, with the increase of $K_{4}$, the system undergoes a series
of topological phase transitions, yielding Floquet HOTPs with more
and more zero and $\pi$ corner modes. The 2D-ORDKL thus provide us
with a nice platform to investigate Floquet HOTPs with multiple corner
states and strong topological signatures. For completeness, we have
also checked the quasienergy spectrum and corner modes in other parameter
regions of the system, and obtain results that are consistent with
the above descriptions.

\begin{figure}
	\begin{centering}
		\includegraphics[scale=0.49]{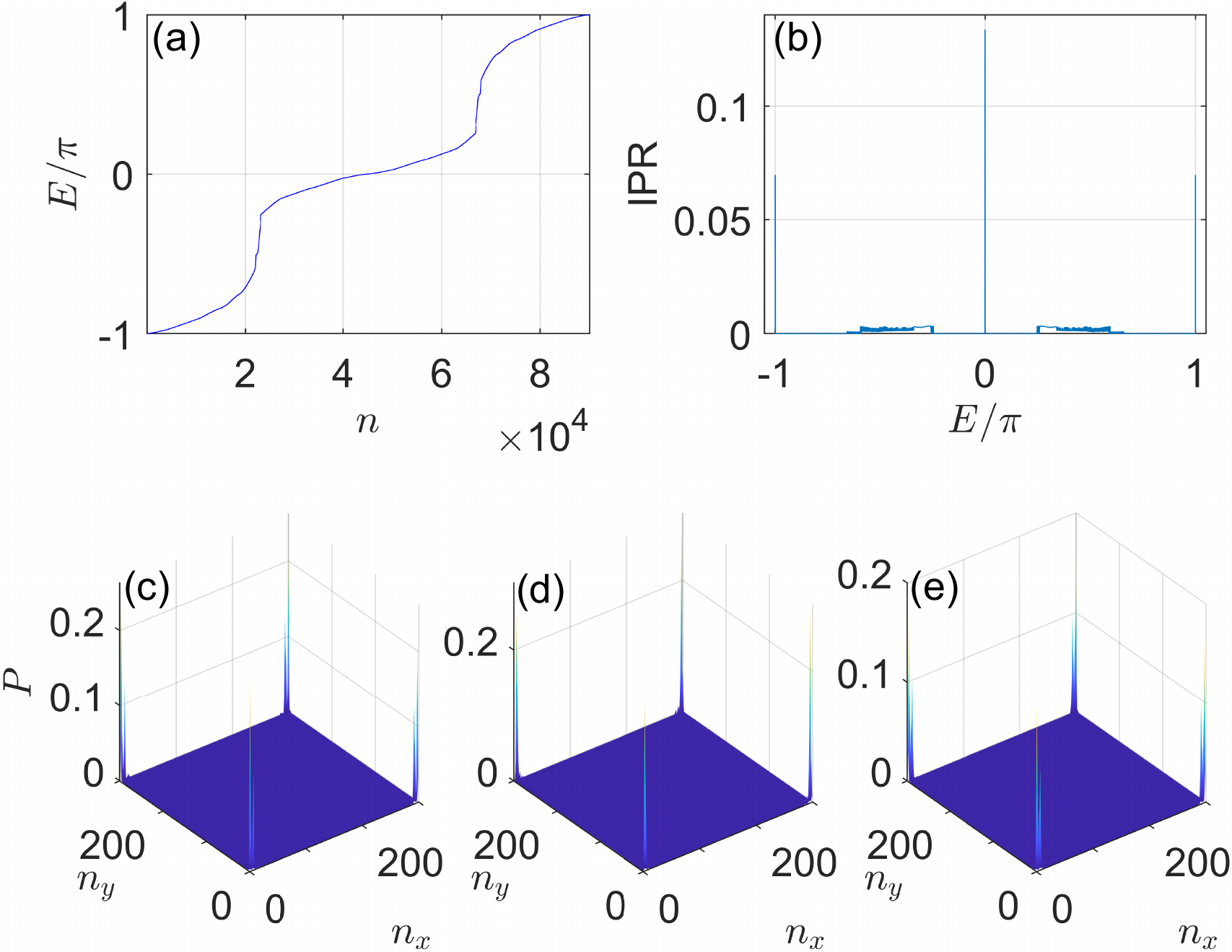}
		\par\end{centering}
	\caption{Quasienergy spectrum, IPR and Floquet zero/$\pi$ corner modes of
		the 2D-ORDKL. System parameters are set as $K_{1}=0.5\pi$, $K_{2}=3.5\pi$,
		$K_{3}=0.5\pi$ and $K_{4}=1.5\pi$. The size of the lattice is $L_{x}=L_{y}=300$.
		In panel (a), $n$ denotes the index of the state. The peaks
		around $E=\pm\pi$ ($E=0$) in panel (b) correspond
		to the IPRs of $\pi$ (zero) Floquet corner modes, whose probability
		distributions are shown in panel (c) {[}(d) and (e){]}.\label{fig:CM}}
\end{figure}

In Fig.~\ref{fig:CM}, we present the quasienergy spectrum, IPR and
corner modes of the system for a typical situation. In Fig.~\ref{fig:CM}(a),
we observe that the Floquet spectrum $E$ spread throughout the first
quasienergy Brillouin zone, and no gaps can be observed around $E=0,\pm\pi$.
However, in Fig.~\ref{fig:CM}(b), we find three clear peaks in the
IPR versus the quasienergy $E$ round $E=0$ and $E=\pm\pi$, which
indicates the existence of localized bound states in the system at
these quasienergies. By investigating the data, we obtain eight (four)
such bound states at $E=0$ ($E=\pm\pi$), with their probability
distributions shown explicitly in Figs.~\ref{fig:CM}(d,e) {[}Fig.~\ref{fig:CM}(f){]}. We see that all of these bound states are indeed
localized around the corners of the system, and their numbers are
determined precisely by the bulk-corner correspondence relation in
Eq.~(\ref{eq:BCC}) for the given set of system parameters. Therefore,
these corner states originate from the higher-order topology of the
2D-ORDKL. They represent topologically degenerate corner modes in
momentum space of the system, which are protected by the chiral symmetry
$\Gamma$. Besides, these corner modes coexist with extended bulk
states at the same quasienergies. We therefore refer to them as corner-localized Floquet
topological bound states in continuum, in the sense that they
do not hybridize with the surrounding bulk states even without a bulk
band gap. This observation also extends the scope of bulk-corner correspondence
of Floquet HOTPs to more general situations, in which the symmetry-protected
corner modes can not only be found in spectral gaps, but also appear
within topological Floquet bands. 
Note in passing that the corner-localized BICs with zero energy have been discovered before in static HOTPs~\cite{BIC1,BIC2,BIC3}.
By contrast, the corner-localized Floquet BICs subject to different topological characterizations, and the corner BICs with quasienergy $E=\pi$ are unique to Floquet HOTPs found in this work.
Experimentally, the implementation
of OBCs in momentum space might be a challenging task. However, by
applying the mapping introduced in Ref.~\cite{ORDKR2} from momentum to position
space lattices, the 2D-ORDKL may also be realized in real space, and
the corner modes may then be probed in setups like photonic waveguide
arrays~\cite{BIC3}. 

\section{Conclusion\label{sec:Sum}}

In this work, we find rich Floquet HOTPs in a 2D extension
of the on-resonance double kicked rotor. Each of the Floquet HOTP
is characterized by a pair of integer topological invariants $(w_{0},w_{\pi})$,
which can be extracted from the dynamics of the system in four distinct
symmetric time frames. The values of $w_{0}$ and $w_{\pi}$ take
quantized jumps whenever the system undergoes a transition between
different Floquet HOTPs. Furthermore, Floquet HOTPs characterized
by arbitrarily large $(w_{0},w_{\pi})$ can be found in principle
with the increase of kicking strengths. Experimentally, the invariants
$(w_{0},w_{\pi})$ could also be obtained by measuring the time-averaged
mean chiral displacements of initially localized wavepackets in different
time frames. Under open boundary conditions, corner-localized modes
with quasienergies zero and $\pi$ are found to be coexisting with
extended bulk states at the same quasienergies, realizing second-order Floquet topological bound
states in continuum. The numbers of these corner modes are further
determined by the bulk topological invariants $(w_{0},w_{\pi})$,
leading to the bulk-corner correspondence of Floquet HOTPs. Even though
the direct detection of these corner modes in momentum space could
be challenging due to the implementation of OBCs,
the double kicked rotor could be mapped to a periodically quenched
tight-binding lattice~\cite{ORDKR2}. Therefore, our model may also be engineered
in two-dimensional photonic or cold atom systems, where the Floquet
corner modes could be imaged in real-space. Putting together, we uncovered
a unique set of Floquet HOTPs in momentum space, which are featured
by large topological invariants and multiple topological corner bound states in continuum. In future
studies, it would be interesting to extend our results to Floquet HOTPs in higher dimensions,
exploring the possibilities of realizing
topological time crystals by superposing these zero and $\pi$ Floquet
corner modes, and investigating the potential applications of these
corner bound states in Floquet quantum computing tasks.


\section*{Acknowledgement}
L.Z. acknowledges Yuhong Zhu for helpful discussions. This work is supported by the National Natural Science Foundation of China (Grant No.~11905211), the China Postdoctoral Science Foundation (Grant No.~2019M662444), the Fundamental Research Funds for the Central Universities (Grant No.~841912009), the Young Talents Project at Ocean University of China (Grant No.~861801013196), and the Applied Research Project of Postdoctoral Fellows in Qingdao (Grant No.~861905040009).

\appendix

\section{Components of the effective Hamiltonian\label{sec:AppA}}
In this appendix, we summarize the explicit expressions of unit vectors
$[n_{\alpha x}(\theta_{x}),n_{\alpha y}(\theta_{x})]$ and $[n_{\beta x}(\theta_{y}),n_{\beta y}(\theta_{y})]$
for $\alpha=1,2$ and $\beta=3,4$ in Eqs.~(\ref{eq:UaThx}) and (\ref{eq:UbThy})
of the main text, respectively. Following the derivation steps as sketched
in the main text and detailed in Refs.~\cite{ORDKR2,ORDKR3} for 1D
descendant models of the 2D-ORDKL, the components of these unit vectors are found to
be
\begin{equation}
n_{1x}(\theta_{x})=-\frac{\cos\frac{\theta_{x}}{2}\sin{\cal K}_{1}\cos{\cal K}_{2}-\sin\frac{\theta_{x}}{2}\sin{\cal K}_{2}}{\sin[E_{x}(\theta_{x})]},\label{eq:n1x}
\end{equation}
\begin{equation}
n_{1y}(\theta_{x})=-\frac{\sin\frac{\theta_{x}}{2}\sin{\cal K}_{1}\cos{\cal K}_{2}+\cos\frac{\theta_{x}}{2}\sin{\cal K}_{2}}{\sin[E_{x}(\theta_{x})]},\label{eq:n1y}
\end{equation}
\begin{equation}
n_{2x}(\theta_{x})=+\frac{\cos\frac{\theta_{x}}{2}\cos{\cal K}_{1}\sin{\cal K}_{2}+\sin\frac{\theta_{x}}{2}\sin{\cal K}_{1}}{\sin[E_{x}(\theta_{x})]},\label{eq:n2x}
\end{equation}
\begin{equation}
n_{2y}(\theta_{x})=+\frac{\sin\frac{\theta_{x}}{2}\cos{\cal K}_{1}\sin{\cal K}_{2}-\cos\frac{\theta_{x}}{2}\sin{\cal K}_{1}}{\sin[E_{x}(\theta_{x})]},\label{eq:n2y}
\end{equation}
and
\begin{equation}
n_{3x}(\theta_{y})=-\frac{\cos\frac{\theta_{y}}{2}\sin{\cal K}_{3}\cos{\cal K}_{4}-\sin\frac{\theta_{y}}{2}\sin{\cal K}_{4}}{\sin[E_{y}(\theta_{y})]},\label{eq:n3x}
\end{equation}
\begin{equation}
n_{3y}(\theta_{y})=-\frac{\sin\frac{\theta_{y}}{2}\sin{\cal K}_{3}\cos{\cal K}_{4}+\cos\frac{\theta_{y}}{2}\sin{\cal K}_{4}}{\sin[E_{y}(\theta_{y})]},\label{eq:n3y}
\end{equation}
\begin{equation}
n_{4x}(\theta_{y})=+\frac{\cos\frac{\theta_{y}}{2}\cos{\cal K}_{3}\sin{\cal K}_{4}+\sin\frac{\theta_{y}}{2}\sin{\cal K}_{3}}{\sin[E_{y}(\theta_{y})]},\label{eq:n4x}
\end{equation}
\begin{equation}
n_{4y}(\theta_{y})=+\frac{\sin\frac{\theta_{y}}{2}\cos{\cal K}_{3}\sin{\cal K}_{4}-\cos\frac{\theta_{y}}{2}\sin{\cal K}_{3}}{\sin[E_{y}(\theta_{y})]}.\label{eq:n4y}
\end{equation}
Here ${\cal K}_{1,2}$ and ${\cal K}_{3,4}$ are given by Eqs.~(\ref{eq:K12})
and (\ref{eq:K34}) of the main text, respectively.


\section{Relation between MCDs and topological invariants\label{sec:AppB}}
In this appendix, we present the explicit relationship between the
time-averaged MCDs and the topological invariants of the 2D-ORDKL
model. Following Eqs.~(\ref{eq:W0PX}), (\ref{eq:W0PY}) and (\ref{eq:MCDW})
in the main text, it is straightforward to verify the following equalities
between the time-averaged MCDs and winding numbers in different combinations
of symmetric frames
\begin{equation}
\overline{C}_{13}+\overline{C}_{14}+\overline{C}_{23}+\overline{C}_{24}=w_{0x}w_{0y},\label{eq:CPPP}
\end{equation}
\begin{equation}
\overline{C}_{13}-\overline{C}_{14}-\overline{C}_{23}+\overline{C}_{24}=w_{\pi x}w_{\pi y},\label{eq:CMMP}
\end{equation}
\begin{equation}
\overline{C}_{13}-\overline{C}_{14}+\overline{C}_{23}-\overline{C}_{24}=w_{0x}w_{\pi y},\label{eq:CMPM}
\end{equation}
\begin{equation}
\overline{C}_{13}+\overline{C}_{14}-\overline{C}_{23}-\overline{C}_{24}=w_{\pi x}w_{0y}.\label{eq:CPMM}
\end{equation}
Combining these equations with the definitions of $w_{0}$ and $w_{\pi}$
in Eqs.~(\ref{eq:W0}) and (\ref{eq:WP}) of the main text, we find
\begin{alignat}{1}
\overline{C}_{0}& \equiv |\overline{C}_{13}+\overline{C}_{14}+\overline{C}_{23}+\overline{C}_{24}|\nonumber \\
& + |\overline{C}_{13}-\overline{C}_{14}-\overline{C}_{23}+\overline{C}_{24}|=w_{0},\label{eq:C0W0}
\end{alignat}
\begin{alignat}{1}
\overline{C}_{\pi}& \equiv |\overline{C}_{13}-\overline{C}_{14}+\overline{C}_{23}-\overline{C}_{24}|\nonumber \\
& + |\overline{C}_{13}+\overline{C}_{14}-\overline{C}_{23}-\overline{C}_{24}|=w_{\pi}.\label{eq:CPWP}
\end{alignat}
Therefore, by measuring the time-averaged MCDs $\overline{C}_{\alpha\beta}$
for $\alpha=1,2$ and $\beta=3,4$, we could obtain the invariants
$(w_{0},w_{\pi})$ of Floquet HOTPs from wavepacket dynamics in different time frames.


\end{document}